%% file: main.tex
\documentclass[conference]{IEEEtran}
\IEEEoverridecommandlockouts

\def\BibTeX{{\rm B\kern-.05em{\sc i\kern-.025em b}\kern-.08em
		T\kern-.1667em\lower.7ex\hbox{E}\kern-.125emX}}
\input{commands}

\allowdisplaybreaks 
\begin{document}
	
	\title{OFDM Based Bistatic Integrated Sensing and Communication: Sensing Beyond CP Limit}
	\thispagestyle{plain}
	\author{Cuneyd Ozturk  \emph{Member, IEEE}, 
		and Cagri Goken \emph{Member, IEEE}
		\thanks{Authors are with Aselsan Inc., Ankara, 06800, Turkey (E-mail: cuneydozturk@aselsan.com, cgoken@aselsan.com)
	}}
	\maketitle
	\begin{abstract}
		This work investigates a bistatic OFDM-based integrated sensing and communication (ISAC) system under a single-target scenario, considering both line-of-sight (LOS) presence and LOS blockage cases. A sliding window-based sensing receiver architecture is proposed to extend the intersymbol interference (ISI)-free sensing range beyond the cyclic prefix (CP) duration by exploiting pilot symbols embedded in the time-frequency grid. The performance of the proposed receiver is evaluated in terms of range and velocity estimation accuracy and is compared against the Cramer-Rao bounds (CRBs) for the bi-static ISAC setting. Numerical results confirm that the proposed method achieves estimation performance that closely approaches the CRBs in the high signal-to-noise ratio (SNR) regime.
	\end{abstract}
	
	\begin{IEEEkeywords} 
		Integrated sensing and communication (ISAC), orthogonal frequency division multiplexing (OFDM), cyclic prefix (CP),  bi-static ISAC, Cramer-Rao bound (CRB), pilot symbols. 
	\end{IEEEkeywords}
	
	\section{Introduction}\label{sec:Intro}
	Integrated sensing and communication (ISAC) systems leveraging OFDM have emerged as a promising paradigm, largely due to the widespread adoption of OFDM in existing wireless communication standards.  This compatibility makes it an attractive and practical solution for ISAC architectures that seek to jointly perform communication and sensing without requiring significant waveform redesign \cite{Gonzalez2024ISAC, Furkan25Holistic, Swindlehurst2024Joint3D, 22_Pucci_SystemLevelAnalysis, Ozturk2025bistatic, bacchielli2024bistaticsensingthzfrequencies, Pucci2022_Bistatic, brunner2024bistaticofdmbasedisacovertheair, Natajara2024BistaticRadar}. ISAC is a promising technology for military communication scenarios, owing to its ability to provide critical sensing capabilities using existing tactical communication infrastructure\cite{Cho24MILCOM, Sharma22MILCOM}. This dual-functionality makes ISAC particularly attractive in tactical environments where the deployment of dedicated and costly radar systems may not be feasible. In particular, ISAC holds significant potential for detecting low-altitude drones, which pose an increasing threat in modern battlefield operations.

	In OFDM-based radar systems, target range and velocity can be estimated efficiently using 2D-FFT based processing at the receiver \cite{Gonzalez2024ISAC, braun2014ofdm, 22_Pucci_SystemLevelAnalysis}. However, a key limitation in many of these studies is the assumption that all target reflections are received within the cyclic prefix (CP) duration to avoid inter-symbol interference (ISI) and preserve OFDM orthogonality \cite{Gonzalez2024ISAC, 22_Pucci_SystemLevelAnalysis, Keskin2021ICIFoe}. This condition restricts the maximum sensing range to $c\Tcp/2$ in mono-static configurations and $c\Tcp$ in multi-static scenarios,  where $c = 3\times 10^8~$[m/s] and $\Tcp$ is the CP duration. For instance, under 5G numerology with a subcarrier spacing of 240 kHz, the corresponding maximum ISI-free range is only 43.5 meters in the mono-static case which is an impractically short distance for many real world applications.

	Recently, there has been a growing interest in extending the sensing range in OFDM-based ISAC networks beyond the conventional CP-limited regime \cite{Tang24CP, Wang25CP, Jiang25CP, xu2025CP}. In \cite{Tang24CP}, the authors demonstrate that by allocating both zero-power and non-zero-power reference signals, the ISI-free range can be extended to the CP duration plus half of the data symbol duration. In \cite{Wang25CP}, the case where the echo delay exceeds the CP duration but remains within the overall OFDM symbol duration is examined. Their approach involves appending a portion of each OFDM symbol to the front through coherent compensation, effectively improving the signal-to-interference-plus-noise ratio (SINR).  In \cite{Jiang25CP}, the authors demonstrate that introducing a timing advance mechanism can effectively extend the sensing range. Moreover, \cite{xu2025CP} proposes a sliding window detection approach that allows for ISI-free sensing beyond the conventional CP-limited range.

	In this manuscript, we aim to extend the ISI-free sensing range in bi-static ISAC systems. To this end, we propose a sliding window detection approach that leverages pilot symbols embedded in the time-frequency grid. The problem is addressed under both line-of-sight (LOS) and LOS-blocked scenarios. The performance of the proposed algorithm is evaluated by comparing it with theoretical bounds for range and velocity estimation, as a function of the pilot pattern, as introduced in \cite{Ozturk2025bistatic}. Simulation results demonstrate that the proposed method achieves the Cramer-Rao bounds (CRBs) for both range and velocity estimation.
	
	This manuscript differs from prior works such as \cite{Tang24CP, Wang25CP, Jiang25CP, xu2025CP} by focusing specifically on bi-static ISAC systems. In contrast to mono-static scenarios where the sensing receiver typically has access to all transmitted modulation symbols, the bi-static configuration presents additional challenges due to the limited availability of this information at the sensing receiver. As a result, the receiver relies solely on pilot symbols to detect the presence of a target and estimate its parameters, namely range and velocity. Notably, the unavailability of all modulation symbols precludes time-domain signal reconstruction and cancellation techniques, such as the one proposed in \cite[Alg.~1]{xu2025CP}. Furthermore, previous works generally assume that the ISAC system operates in full-duplex mode with perfect self-interference cancellation. In the bi-static context, this issue is mirrored by the presence of a  LOS path between the transmitter and receiver that is explicitly addressed in our work. Extension of the proposed approach to multi-static ISAC architectures is left as a direction for future research.
	\vspace{-0.3cm}
	\section{System Model}\label{sec:SystemModel}
	In this section, we describe the signal model and formally define the problem. We consider a bi-static configuration based on OFDM signaling, in which both the transmitter and the receiver are equipped with multiple antennas. A single target is assumed to be present in the environment.  Importantly, the propagation delay of the reflected signal is not restricted to be shorter than the CP duration. A portion of the modulation symbols is allocated to pilot symbols, which are known at the sensing receiver. By leveraging these pilot symbols, the sensing receiver aims to detect the presence of the target and estimate its range and velocity. We consider two different scenarios based on presence of the LOS path as illustrated in Figs.~\ref{fig:ScenarioI}-\ref{fig:ScenarioII}.
	\subsection{Transmitted Signal}
	The transmitter employs OFDM signaling with carrier frequency $f_c$.  $\Nsc$ and $\Msym$ denote the number of subcarriers and OFDM symbols in a single OFDM frame, respectively. The subcarrier spacing is denoted by $\Delta f$,  and the duration of a single OFDM symbol, including the CP, is given by $\Tsym = \Tcp + \Td$, where $\Tcp$ is the CP duration and $\Td = 1/\Delta f$ is the data symbol duration.  Consequently, the sampling period is $\Ts  = \Td/\Nsc$. The number of CP samples per OFDM symbol is given by $\Ncp = \lceil \Tcp/\Ts \rfloor$, $\lceil \cdot \rfloor$ denotes rounding to the nearest integer. Thus, the total number of samples per OFDM symbol is $\Nsym = \Nsc + \Ncp$.
	
	The baseband transmitted OFDM signal is denoted by $s(t)$ and defined as
	\begin{align}
		s(t) = \frac{1}{\sqrt{\Nsc}} \sum_{m=0}^{\Msym-1} \sum_{n=0}^{\Nsc-1} X[m, n] e^{j 2\pi n \Delta f   t} u(t-m\Tsym),
	\end{align}
	where
	\begin{align}
		u(t) = 
		\begin{cases}
			1, &~\text{if}~ 0\le t < \Tsym, \\
			0, &~\text{otherwise.}
		\end{cases}
	\end{align}
	and $X[m, n]$ denotes the modulation symbol corresponding to the $m^{\text{th}}$ OFDM symbol and the     $n^{\text{th}}$ subcarrier. Throughout this manuscript, QPSK modulation is assumed, i.e., $\abs{X[m, n]} = 1$ for all $m, n$. 
	
	Let $\mathcal{P}$ denote the set of pilot symbols in the time-frequency grid; that is, if $(m, n)\in\mathcal{P}$, then $X[m, n]$ is a pilot symbol known to the sensing receiver. Throughout this manuscript, a periodic pilot pattern is assumed. Specifically,  the pilot symbol set is defined as
	\begin{align}
		\mathcal{P} \triangleq \left\{(m, n) \Bigg\lvert~\frac{n}{n_p}\in\mathbb{N},~\text{and}~\frac{m}{m_p}\in\mathbb{N}\right\}.
	\end{align}
	where $n_p$ and $m_p$ denote the pilot spacing in frequency and time, respectively. The total number of pilot symbols is given by 
	\begin{align}
		\abs{\mathcal{P}} = \left(\Big\lfloor \frac{\Nsc-1}{n_p} \Big\rfloor + 1\right)  \left(\Big\lfloor \frac{\Msym-1}{m_p} \Big\rfloor + 1\right),
	\end{align}
	where $\abs{\cdot}$ denotes the cardinality of a set.  The ratio of the total time-frequency resources allocated to the pilot symbols is defined as $0\le \rho \triangleq \abs{\mathcal{P}}/(\Nsc \Msym) \le  1$.
	
	Let $N_T$ denote the number of antenna elements at the transmitter. It is assumed that the transmitter employs a beamformer $\fbt\in\mathbb{C}^{N_T\times 1}$, under the assumption that a single data stream is transmitted. The transmitted passband signal for a single OFDM frame is thus given by $\Re\{\fbt s(t) e^{j 2 \pi f_c t}\}$.
	
	\vspace{-0.3cm}
	\subsection{Sensing Channel and Received Signal}
	
	\begin{figure}
		\centering
		\includegraphics[width = 0.7\linewidth]{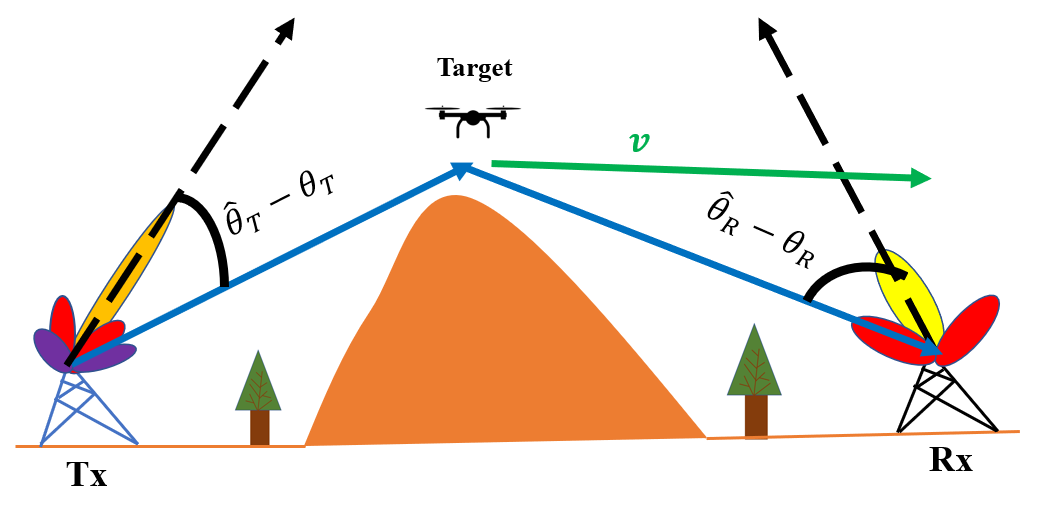}
		\caption{System geometry when the LOS path between the sensing transmitter and the sensing receiver is blocked (Scenario I).}
		\label{fig:ScenarioI}
		\vspace{-0.2cm}
	\end{figure}
	
	\begin{figure}
		\centering
		\includegraphics[width =0.7\linewidth]{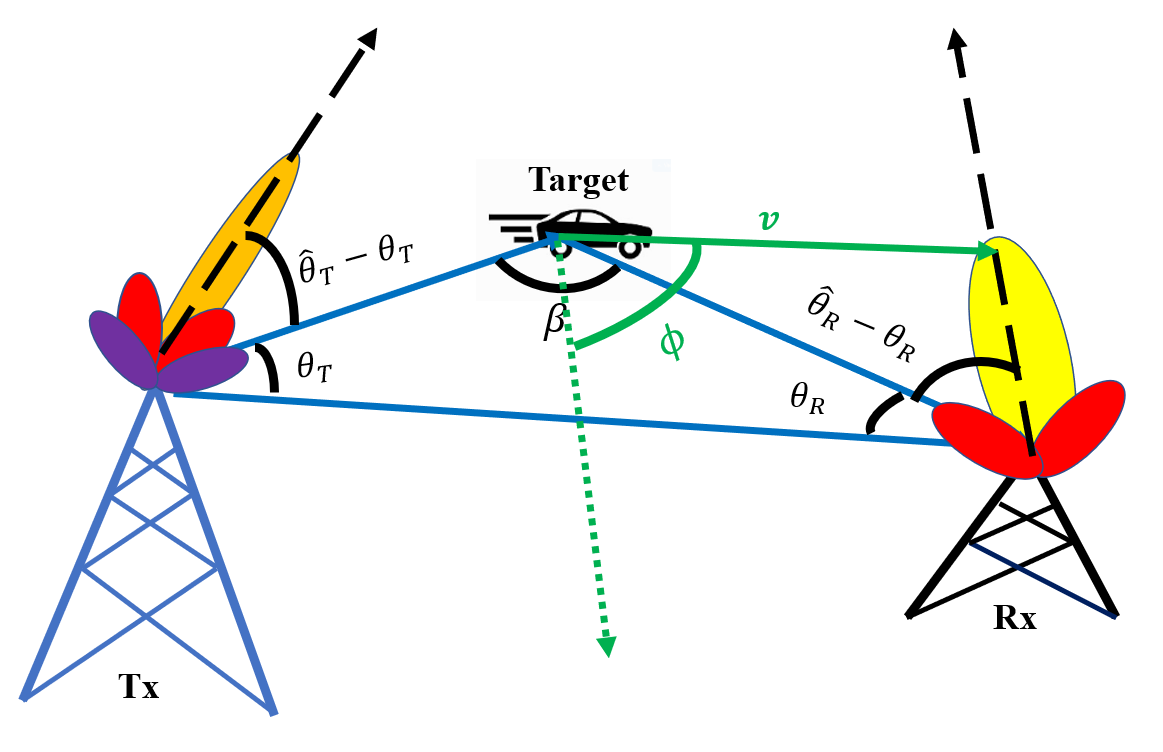}
		\caption{System geometry when the LOS path between the sensing transmitter and the sensing receiver is present (Scenario II).}
		\label{fig:ScenarioII}
		\vspace{-0.5cm}
	\end{figure}
	
	The transmitter and the receiver are located at positions $\pbtx$ and $\pbrx$, respectively. It is assumed that the transmitter and the receiver are stationary. A single target is considered, located at position $\pb$ with velocity $\vb$. The bi-static range is given by
	\begin{align}
		\Rbis = \norm{\pbtx-\pb} + \norm{\pb-\pbrx},
	\end{align}
	and the corresponding propagation delay for the reflected path is
	\begin{align}
		\tau_{\text{NLOS}} = \frac{\Rbis}{c}.
	\end{align} 
	In addition,  the propagation delay between for the LOS path is expressed as
	\begin{align}
		\tau_{\text{LOS}} = \frac{\norm{\pbtx-\pbrx}}{c}.
	\end{align}
	The Doppler shift between the target and the sensing receiver is expressed as \cite{willis2005bistatic}
	\begin{align} 
		f_D = \frac{2 \norm{\vb}}{\lambda} \cos\phi \cos\left(\frac{\beta}{2}\right),
	\end{align}
	where $\lambda = c/f_c$ is the wavelength, $\beta$ is the bi-static angle and $\phi$ is the angle between the target's velocity vector and the bisector of the bi-static angle as depicted in Fig.~\ref{fig:ScenarioII}. The bi-static angle $\beta$ is computed as
	\begin{align}
		\beta = \cos^{-1}\left( \frac{\norm{\pbtx-\pb}^2 + \norm{\pb-\pbrx}^2-\norm{\pbtx-\pbrx}^2}{2 \norm{\pbtx-\pb} \norm{\pb-\pb}}\right).
	\end{align}
	The angle $\phi$ is unknown, only the projected velocity $\vbis\triangleq \norm{\vb} \cos\phi$ can be estimated at the receiver.

	Moreover, the receiver employs a beamformer $\fbr\in\mathbb{C}^{N_R\times 1}$ where $N_R$ is the number of receive antenna elements.  By denoting $\theta_T$ and $\theta_R$ as the angle-of-departure (AoD) and angle-of-arrival (AoA) corresponding to the reflected path, multiplication of the steering vectors are denoted as
	\begin{align}
		\Wb = \boldsymbol{a}_{R}(\theta_R)\boldsymbol{a}_{T}(\theta_T)^{\hermit}  \in\mathbb{C}^{N_R\times N_T},
	\end{align}
	where $[\boldsymbol{a}_{T}(\theta)]_\ell = \exp\left(j \pi \ell \sin\theta \right) \sqrt{1/N_R}$ and $[\boldsymbol{a}_{R}(\theta)]_\ell = \exp\left(j \pi \ell \sin\theta \right) \sqrt{1/N_T}$ by assuming the antenna element spacing is $\lambda/2$ for both transmit and receive antenna arrays.

	It is important to note that mismatches in antenna orientations can be incorporated into the constant term $\fbr^\hermit \Wb \fbt$,  which captures the combined effects of beamforming gains at the transmitter and receiver. Specifically, it is assumed that $\fbt = \ab_T(\widehat{\theta}_T)$ and  $\fbr = \ab_R(\widehat{\theta}_R)$ for some $\widehat{\theta}_T$ and $\widehat{\theta}_R$, which may deviate from their true values $\theta_T$ and $\theta_R$ as depicted in Figs.~\ref{fig:ScenarioI}-\ref{fig:ScenarioII}.
	
	We consider two scenarios based on the presence or absence of a LOS path. The received signal in each scenario is modeled as follows:
	
	\begin{enumerate}
		\item \textit{Scenario I: LOS is blocked}  
		\begin{align}
			r(t) &= \alpha_{\text{NLOS}} s(t-\tau_{\text{NLOS}}) e^{j 2 \pi f_D t}  + z(t),
		\end{align}
		\item \textit{Scenario II: LOS is present}
		\begin{align}
			r(t) &= \alpha_{\text{LOS}} s(t-\tau_{\text{LOS}}) + \alpha_{\text{NLOS}} s(t-\tau_{\text{NLOS}}) e^{j 2 \pi f_D t} \nonumber \\
			& + z(t),
		\end{align}
	\end{enumerate}
	where $\alpha_{\text{NLOS}}\in\mathbb{C}$ and $\alpha_{\text{LOS}}\in\mathbb{C}$ denote the channel gains associated with the LoS and NLOS components, respectively, and $z(t)$ represents the additive Gaussian noise with power spectral density level as $N_0$ at the sensing receiver. 
	Since the term $\fbr^\hermit \Wb \fbt$ is constant  across both scenarios, it can be absorbed into the signal and noise scaling can be done accordingly, allowing for normalization in the expressions below \cite{Pucci2022_Bistatic}
	\begin{align}\label{eq:alphaNLOS}
		\alpha_{\text{NLOS}} = \frac{ e^{-j 2\pi f_c \tau_{\text{NLOS}}}\lambda \sqrt{\sigma_{\text{RCS}} }} {(4\pi)^{3/2} \norm{\pbtx-\pb} \norm{\pb-\pbrx}}
	\end{align}
	and
	\begin{align} \label{eq:alphaLOS}
		\alpha_{\text{LOS}} = \frac{e^{-j 2\pi f_c \tau_{\text{LOS}}}\lambda} {4\pi \norm{\pbtx-\pbrx}}
	\end{align}
	where $\sigma_{\text{RCS}}$ is the radar-cross section.

	\subsection{Problem Definition}
	
	In both scenarios (Scenario I-II), the sensing receiver aims to detect presence of the target, and if so, proceeds to estimate the target's range and velocity based on the received samples $\{r[k]\}_k$ where $r[k] \triangleq r(k\Ts)$ for  $0\le k \le \Msym \Nsym-1$.  In both considered scenarios, the propagation delay is not assumed to be shorter than the cyclic prefix duration $\Tcp$.
	
	In Scenario I, where only the NLOS path is present, the sensing receiver detects the presence of a target by performing hypothesis testing using a sliding window approach applied to the received samples. In Scenario II, which involves both LOS and NLOS components, the receiver first attempts to detect the presence of the LOS path using a similar sliding window-based hypothesis testing method. If the LOS component is detected, it is subsequently removed to enable accurate estimation of the NLOS component. The complete methodologies for both scenarios are detailed in Sec.~\ref{sec:Range_vel_est}.
	
	\vspace{-0.2cm}

	\section{Range and Velocity Estimation}\label{sec:Range_vel_est}
	In this section, we begin by considering the case where the LOS path is blocked (Scenario I). The range and velocity estimation procedure for Scenario I is detailed in Secs.~\ref{subsec:def_hypo}, \ref{subsec:decision_metric}, \ref{subsec:stopping}, and \ref{subsec:threshold}. Subsequently, building upon the methodology developed for Scenario I, we present the extension to Scenario II, where both LOS and NLOS components are present.

	\vspace{-0.2cm}
	\subsection{Definition of the Hypotheses (Scenario I)} \label{subsec:def_hypo}
	
	The propagation delay associated with the NLOS path, denoted by $\tau_{\text{NLOS}}$ can be decomposed with respect to the CP duration $\Tcp$ as follows:
	\begin{align}
		\tau_{\text{NLOS}} = \Tcp \Big \lfloor  \frac{\tau_{\text{NLOS}}}{\Tcp}\Big\rfloor + \epsilon, 
	\end{align}
	where $0\le \epsilon < \Tcp$.  Based on this, the maximum number of CP-length blocks that can fit within the maximum sensing range $\Rmax$\footnote{It should be noted that our approach is not inherently constrained by $\Rmax$, rather, $\Rmax$ is introduced solely to reduce computational complexity.} is defined as
	\begin{align}
		L  \triangleq \Big\lceil \frac{\Rmax}{c \Tcp} \Big\rceil.
	\end{align}
	Accordingly, the set of hypotheses is defined as:
	\begin{align}
		\mathcal{H}_0:&~\text{No target}, \nonumber\\
		\mathcal{H}_\ell:&~(\ell-1)\Tcp \le \tau_{\text{NLOS}} < \ell\Tcp. \label{eq:hypothesis}
	\end{align}
	Under each hypothesis, least squares (LS) channel estimates are computed, and the peak value of the corresponding two-dimensional periodogram is used as the decision metric. The details of this estimation and decision-making procedure are described in the following subsections.
	
	\subsection{Decision Metric Computation  (Scenario I)}\label{subsec:decision_metric}
	Under the hypothesis $\mathcal{H}_\ell$,  the number of received OFDM symbols available within a single OFDM frame is given by\footnote{In this study, the sliding window is restricted to the duration of a single OFDM frame.}
	\begin{align}
		M_\ell \triangleq \Big \lfloor \frac{\Msym \Tsym- (\ell-1)\Tcp}{\Tsym} \Big \rfloor.
	\end{align}
	For each of these $M_\ell$ OFDM symbols, the CP is removed and an $\Nsc$ point DFT is computed.   In particular, for $0\le m \le M_\ell-1$ and $0\le n \le \Nsc-1$, we define
	\begin{align}
		R_\ell[m, n] \triangleq \frac{1}{\sqrt{\Nsc}} \sum_{k= 0}^{\Nsc-1} &r[m \Nsym + \ell\Ncp + k] e^{-j \frac{ 2\pi(k-\Ncp)n}{\Nsc}}. \label{eq:DFT_ell}
	\end{align}
	For any pilot symbol located at $(m, n)\in\mathcal{P}$ with $m\le M_\ell-1$, the least squares (LS) channel estimate is obtained as
	\begin{align}
		\HLS_\ell[m, n] \triangleq R_\ell[m, n] \left(X[m, n]\right)^{*}\label{eq:LS_ell}.
	\end{align}
	Assuming $m_p$ and $n_p$ are the periods of the pilot symbols across time and frequency, the channel       estimates at the pilot grid are collected as
	\begin{align}
		\widehat{h}^{\text{LS}}_\ell[\mu, \nu] \triangleq \HLS_\ell[\mu m_p, \nu n_p]
	\end{align}
	for $0\le \mu \le \lfloor (M_\ell-1)/m_p \rfloor$ and $0\le \nu \le \lfloor (\Nsc-1)/n_p\rfloor$.
	
	Let $M_{\text{per}}$ and $N_{\text{per}}$ denote the 2D FFT sizes across time and frequency, respectively. The corresponding two-dimensional periodogram is computed as
	\begin{align}
		P_\ell(p, q) \triangleq \Bigg\lvert \sum_{\nu = 0}^{N_{\text{per}}-1}  \left(\sum_{\mu = 0}^{M_{\text{per}}-1} 	\widehat{h}^{\text{LS}}_\ell[\mu, \nu] e^{-j 2\pi \mu p/M_{\text{per}}} \right) e^{j 2\pi \nu q/N_{\text{per}}} \Bigg\rvert^2, \label{eq:perio_ell}
	\end{align}
	where $p\in\{-M_{\text{per}}/2, -M_{\text{per}}/2 + 1, \ldots,  M_{\text{per}}/2 - 1\}$ and  $q\in\{0, 1, \ldots,  N_{\text{per}} - 1\}$.
	
	The decision metric for hypothesis for hypothesis $\mathcal{H}_\ell$ is then defined as the peak value of the periodogram:
	\begin{align}
		\eta_\ell \triangleq \max_{(p, q)} 	P_\ell(p, q). \label{eq:decision_metric_ell}
	\end{align}
	
	It should be noted that the methodology described by equations \eqref{eq:hypothesis}, \eqref{eq:DFT_ell}, \eqref{eq:LS_ell}, \eqref{eq:perio_ell}, and \eqref{eq:decision_metric_ell} assumes that the starting sample index is a multiple of $\Ncp$. However, the same computations can be extended to arbitrary sampling indices $k$. Next, we describe the procedure for target detection and joint range and velocity estimation for a given decision threshold $\kappa$, followed by a discussion on how this threshold can be appropriately selected.
	
	\subsection{Stopping Criteria and Thresholding  (Scenario I)}\label{subsec:stopping}

	The detection process begins by evaluating the test statistics $\eta_\ell$ for $1\le \ell \le L$. If $\eta_\ell < \kappa$ for all $\ell$, we select the null hypothesis $\mathcal{H}_0$, declaring that no target is present. Otherwise, let $h$ be the smallest index such that $\eta_h\ge \kappa$. In this case, we declare the presence of a target. To refine the target parameter estimates, we define a finer observation window given by $(h-1)\Tcp \leq t < (h+W-1)\Tcp$, where $W\in\mathbb{N}$ is a design parameter.

	The rationale for introducing such a window is as follows: suppose $\mathcal{H}_{\ell_0}$ is the true hypothesis. As $\ell$ approaches $\ell_0$, the correlation between $\eta_\ell$ and $\eta_{\ell_0}$ increases. Consequently, in the presence of a target, it is likely that $\eta_\ell$ may attain large values  even when $\ell \neq \ell_0$ provided that $\ell$ is close to $\ell_0$. Therefore, selecting the first index exceeding the threshold may lead to an incorrect hypothesis.
	
	Next, we compute the maximums of the periodograms assuming the propagation delay lies within  $[k\Ts, k\Ts + \Tcp)$ for all sampling indices $k\in \mathcal{W}\triangleq \left\{(h-1)\Ncp, \ldots, (h+W-1)\Ncp-1\right\}$ by following  methodology described in Sec.~\ref{subsec:decision_metric}. Among these, the index yielding the maximum peak is selected and denoted by $\widehat{k}$. In particular, our final decision is that the propagation delay is between $\widehat{k}\Ts$ and $\widehat{k}\Ts + \Tcp$. Let $\widehat{P}(p, q)$ denote the corresponding two-dimensional periodogram. The range and velocity estimates are obtained as follows:
	\begin{enumerate}
		\item First, maximizing indices of the periodogram is found as $(\widehat{p}, \widehat{q}) = \argmax \widehat{P}(p, q)$.
		\item Second, quadratic interpolation is applied following the method described in \cite{braun2014ofdm}, allowing for refined estimation of the peak location. Let $\widetilde{p}$ and $\widetilde{q}$ denote the interpolated frequency and time indices, respectively, which may include fractional components.
		\item The range estimate is expressed as
		\begin{align}
			\Rhatbis = \frac{\widetilde{p} c}{n_p \Delta f N_p}. \label{eq:R_hat_bis}
		\end{align}
		\item By using $\widehat{\theta}_R$, the distances of the target to the receiver and the transmitter can be estimated as 
		\begin{align}
			\widehat{d}_{\text{rx}} = \frac{\Rbis^2-\norm{\pbtx-\pbrx}^2}{2\left( \Rbis-\norm{\pbtx-\pbrx} \cos \widehat{\theta}_R \right)},~ \widehat{d}_{\text{tx}} = \Rbis-\widehat{d}_{\text{rx}}. \label{eq:drx_hat}
		\end{align}
		\item The bi-static angle is estimated as
		\begin{align}
			\widehat{\beta} = \cos^{-1}\left( \frac{\widehat{d}_{\text{tx}}^2 + \widehat{d}_{\text{rx}}^2-\norm{\pbtx-\pbrx}^2 }{2 \widehat{d}_{\text{tx}}\widehat{d}_{\text{rx}}}\right). \label{eq:beta_est}
		\end{align}
		\item Finally, the bi-static velocity can be esimated as
		\begin{align}
			\widehat{v}_{\text{bis}} = \frac{\widetilde{q}c}{2f_c m_p \Tsym M_p \cos\left(\widehat{\beta}/2\right)}. \label{eq:v_hat_bis}
		\end{align}
	\end{enumerate}
	
	As can be seen from equations \eqref{eq:drx_hat}, \eqref{eq:beta_est}, and \eqref{eq:v_hat_bis}, the estimation of bi-static velocity $\widehat{v}_{\text{bis}}$ requires knowledge of the distance between the transmitter and the receiver at the sensing receiver. It is also important to highlight that the impact of AoA mismatch affects range and velocity estimations differently. In the high SNR regime, the bi-static range $\Rbis$ can still be estimated accurately despite AoA mismatches. However, due to its explicit dependence on directional information as indicated in \eqref{eq:drx_hat}, $\vbis$ cannot be perfectly recovered in the presence of AoA error, even in a noiseless scenario.
	
	\vspace{-0.2cm}
	\subsection{Threshold Computation (Scenario I)} \label{subsec:threshold}
	Let $\ell_0$ denote the true hypothesis, meaning the bi-static range satisfies $(\ell_0-1)\Tcp \le         \Rbis < \ell_0 \Tcp$. Then, given windowing size $W$, we define the complementary window of interest    as
	\begin{align}
		\mathcal{W}^c_0 \triangleq \{1, \ldots, L\} \setminus \{\ell_0-\lfloor W/2 \rfloor, \ldots, \ell_0 + \lfloor (W+1)/2 \rfloor \}.
	\end{align}
	which excludes the region surrounding the true hypothesis.
	
	The index corresponding to the maximum decision metric outside the window of interest is then defined as
	\begin{align}
		\widehat{\ell} = \argmax_{\ell \in \mathcal{W}^c_0 } \eta_{\ell}
	\end{align} 
	The threshold $\kappa$ is subsequently determined such that the probability of a false alarm           satisfies $\Pr\{\eta_{\widehat{\ell}} \ge \kappa \} = P_f$, where $P_f$ is the desired false-alarm probability.
	\vspace{-0.2cm}
	\subsection{Extension to Scenario II}\label{subsec:ScenarioII_range_vel_est}
	
	When the LOS path is present, our first objective is to detect this component. In this section, we assume that the delay difference between the NLOS and LOS paths satisfies $\tau_{\text{NLOS}}-\tau_{\text{LOS}}\le \Tcp$. Under this assumption, we again consider a set of $L+1$ hypotheses, where $\mathcal{H}_\ell$ corresponds to $(\ell-1)\Tcp \le \tau_{\text{LOS}} \le \ell \Tcp$ for $\ell \ge 1$ and $\mathcal{H}_0$ denotes the absence of an LOS path. For LOS detection and range estimation, we follow the same procedures as described in Secs.~\ref{subsec:decision_metric} and \ref{subsec:threshold}.
	
	If the LOS path is detected, let $\widehat{\tau}_{\text{LOS}}$ denote the estimated LOS propagation delay. Using this estimate, the complex path gain $\alpha_{\text{LOS}}$ can be estimated from \eqref{eq:alphaLOS} as $\widehat{\alpha}_{\text{LOS}}  = e^{-j 2\pi f_c \widehat{\tau}_{\text{LOS}}}\lambda /(\pi\widehat{\tau}_{\text{LOS}} c)$.
	Using this estimate, the LOS channel contribution across the time-frequency grid can be synthesized.  Let $w = \lfloor \tau_{\text{LOS}}/\Ts \rfloor$ denote the initial sampling index corresponding to the LOS path. The LS channel estimates computed at this index are denoted as $\HLS_{w}$.  The estimated LOS component is subtracted from each pilot symbol according to: $\HLS_w[m,n] - \widehat{\alpha}_{\text{LOS}} e^{-j 2\pi n \Delta f \widehat{\tau}_{\text{LOS}}} \quad \text{for } (m,n) \in \mathcal{P}.$
	Following the removal of the LOS component, the corresponding periodogram is computed, and the bi-static range and velocity are estimated using the procedure outlined in equations~\eqref{eq:R_hat_bis}-\eqref{eq:v_hat_bis}. 
	

	\section{Numerical Results}

	In this section, we present numerical results to validate the effectiveness of the proposed approach. We begin with Scenario I, where only the NLOS path is present, and demonstrate the range and velocity estimation performance under this condition. Subsequently, we introduce the LOS path and evaluate the performance of the proposed method in terms of range and velocity estimation in the presence of both LOS and NLOS components.
	
	In the system setup, the target location $\pb = [x, y]$ is randomly generated with $x\sim \mathcal{U}[x_{\min}, x_{\max}]$ and $y\sim \mathcal{U}[y_{\min}, y_{\max}]$. The target velocity magnitude is drawn from $\norm{\vb}\sim \mathcal{U}[v_{\min}, v_{\max}]$, and $\phi$ is sampled from $\mathcal{U} [\phi_{\min}, \phi_{\max}]$. The noise variance is computed as $\sigma^2 = N_0 \Nsc \Delta f N_F$, where $N_F$ is the noise figure. The system parameters are summarized in Tab.~\ref{tab:system_params}. The CP duration is set to 1~[$\mu$s], which corresponds to an ISI-free sensing range of $300~$ meters. The maximum allowable sensing distance is set to $R_{\max} = 3000$ meters,  implying that a maximum of 10 CP-length blocks can fit within the total propagation delay.

	\begin{table}
		\footnotesize
		\begin{center}
			\begin{tabular}{| l ||c |} 
				\hline
				$f_c$ (Carrier frequency) & 30~ [GHz] \\
				\hline
				$\Delta f$ (Subcarrier spacing) &    200~[kHz]\\
				\hline
				$\Tcp$ (CP duration) &    1~[$\mu$s]\\
				\hline
				$\Nsc$ ($\#$ of subcarriers) & 70 \\
				\hline
				$\Msym$ ($\#$ of symbols) & 100 \\
				\hline
				$N_{\text{per}}$ (FFT size over subcarriers) &$2^{10}$\\
				\hline
				$M_{\text{per}}$ (FFT size over symbols) & $2^{10}$\\
				\hline
				$N_0$ (Noise power-spectral density) & $-174~$dBm/Hz \\
				\hline
				$N_F$ (Noise figure) & 8 dB\\
				\hline 
				$[v_{\min}, v_{\max}]$ & [0, 30]~[m/s]\\
				\hline
				$[\phi_{\min}, \phi_{\max}]$ & [-5$^\circ$, 5$^\circ$] \\
				\hline
			\end{tabular}
		\end{center}
		\caption{\scriptsize{System Parameters}}
		\label{tab:system_params}
		\vspace{-0.7cm}
	\end{table}
	
\begin{figure}
		\centering
		\includegraphics[width=0.65\linewidth]{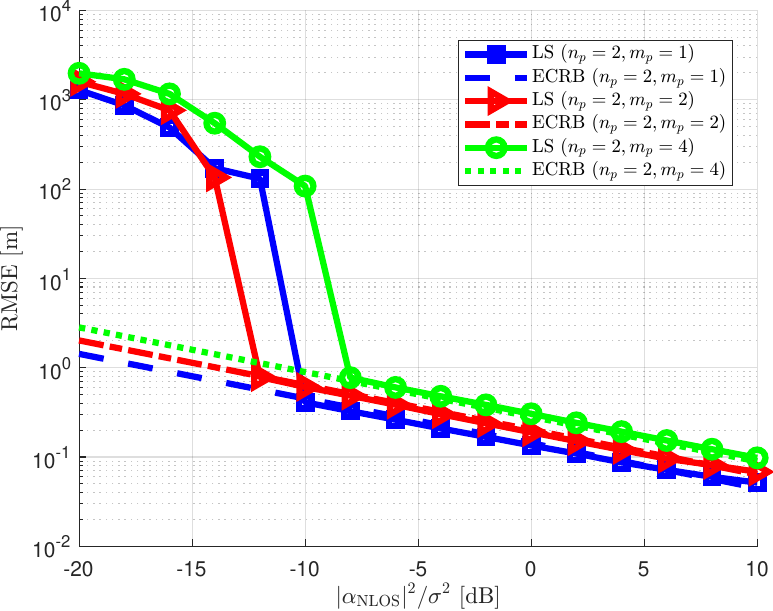}
		\caption{RMSE of the bi-static range estimation along with the ECRB values when the LOS path is blocked. }
		\label{fig:range_RMSE}
		\vspace{-0.3cm}
	\end{figure}

\begin{figure}
		\centering
		\includegraphics[width=0.65\linewidth]{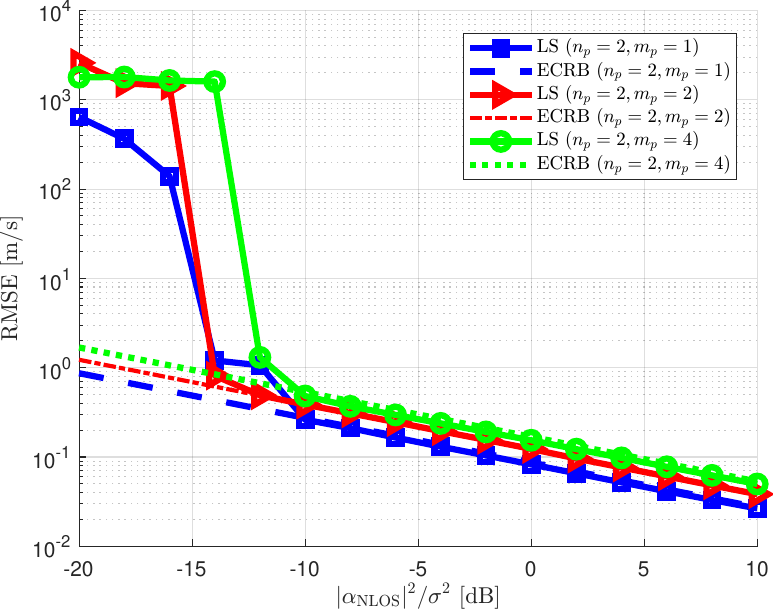}
		\caption{RMSE of the bi-static velocity estimation along with the ECRB values when the LOS path is blocked.}
		\label{fig:velocity_RMSE}
		\vspace{-0.3cm}
	\end{figure}

	In Figs.~\ref{fig:range_RMSE}-\ref{fig:velocity_RMSE}, we present the RMSEs of range and velocity estimations obtained using our proposed approach based on LS channel estimates, under Scenario I. The RMSEs are evaluated as a function of the signal-to-noise ratio defined by $\abs{\alpha_{\text{NLOS}}}^2/\sigma^2$, and are computed over multiple realizations of target location, velocity, and noise, conditioned on successful target detection. For the velocity estimation results shown in Fig.~\ref{fig:velocity_RMSE}, it is assumed that the angle of AoA is perfectly known at the sensing receiver.

	Three different periodic pilot patterns are considered, corresponding to $(n_p, m_p)\in\{(2, 4),~ (2, 2), ~ (2, 1)\}$. The associated pilot overhead ratios are $\rho\in\{0.125, 0.25, 0.5\}$, respectively. The transmitter and receiver are positioned at $[-1000,~ 0]$~[m] and $[1000,~ 0]$~[m], respectively.  The target location is randomly generated within the region defined  $x_{\min} = y_{\min} =  -1000$ [m],  and $x_{\max} = 1000$ [m], $y_{\max} = -500$ [m]. It is worth noting that, under this configuration, the bi-static range is strictly greater than 2000 meters, resulting in a round-trip propagation delay that spans at least the 7\textsuperscript{th} CP block. For the thresholding step in the sliding window-based detection algorithm (see Secs.~\ref{subsec:stopping}-\ref{subsec:threshold}), the window size is set to $W = 2$, and the false alarm probability is configured as $P_f = 10^{-3}$.
	
	To benchmark the performance of the proposed approach, we utilize the CRB expressions for range and velocity estimation derived in \cite{Ozturk2025bistatic}. Although these CRBs are obtained under the assumption that the NLOS propagation delay satisfies  $\tau_{\text{NLOS}}\le \Tcp$, they remain valid as exact theoretical bounds when conditioned on correctly identifying the true hypothesis found via the procedures outlined in Secs.~\ref{subsec:decision_metric}, \ref{subsec:stopping}, and \ref{subsec:threshold}. This is confirmed by the simulation results shown in Figs.~\ref{fig:range_RMSE}-\ref{fig:velocity_RMSE}. When plotting the CRB curves, we compute their ensemble averages over randomly generated target positions to obtain the expected CRBs (ECRBs), and we display their square roots to facilitate comparison with RMSE values. The results indicate that the proposed algorithm closely follows the CRB trend in the high-SNR regime. Furthermore, it is observed that increasing the pilot overhead ratio from  $0.125$ to $0.5$ yields only marginal performance gains in both range and velocity estimation, suggesting that relatively sparse pilot patterns may be sufficient for accurate sensing in the considered system setting.
	
	Additionally, the impact of AoA errors on velocity estimation performance is quantified in Tab.~\ref{tab:AoA mismatch} for pilot patterns  $(n_p, m_p) = (2, 1), (2, 2)$  at an SNRs of 0 and 10 dB. The results demonstrate that inaccuracies in AoA estimation can significantly degrade velocity estimation accuracy. Therefore, to ensure robust performance, accurate estimation of AoA should be incorporated into the sensing receiver design.
	
	\begin{table}
		\begin{center}
			\footnotesize
			\begin{tabular}{ c ||c  || c } 
				SNR [dB]  & $\abs{\widehat{\theta}_R-\theta_R}$ &  RMSE [m/s]  \\
				\hline
				0	  &   $0^\circ$ & 0.0826\\
				\hline
				0  &   $1^\circ$ & 0.1471\\
				\hline
				0  &   $5^\circ$ & 0.5635\\
				\hline
				10 &   $0^\circ$ & 0.0265\\
				\hline
				10  &   $1^\circ$ & 0.1315\\
				\hline
				10 &   $5^\circ$ & 0.5620\\
				\hline
			\end{tabular}
		\end{center}
		\caption{\scriptsize{RMSE [m/s] for various values of AoA error and $(n_p, m_p) = (2, 1)$ when SNR $\in\{0, 10\}~$ \MakeLowercase{dB}.}}
		\label{tab:AoA mismatch}
		\vspace{-0.7cm}
	\end{table}
	
\begin{figure}
		\centering
		\includegraphics[width=0.65\linewidth]{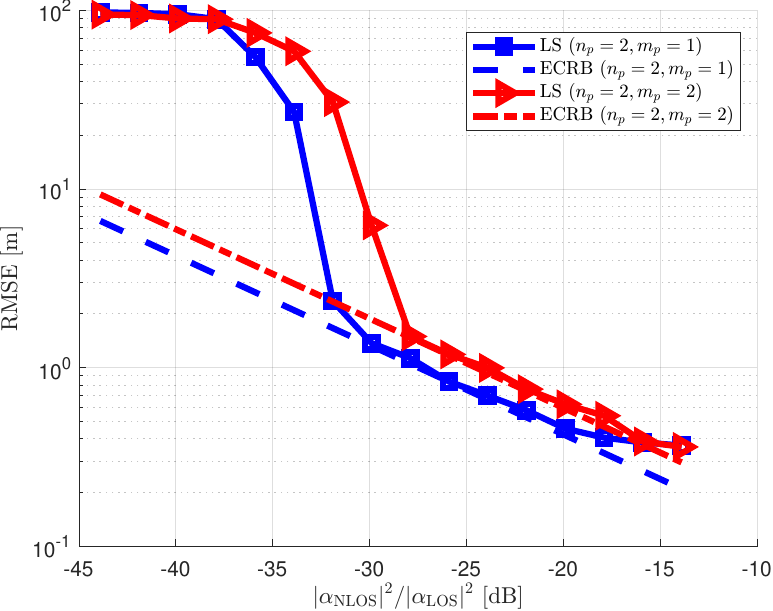}
		\caption{RMSE of the bistatic range estimation under LOS path along with the ECRB values.}
		\label{fig:range_RMSE_NLOS_LOS}
		\vspace{-0.5cm}
	\end{figure}

	Finally, in Fig.~\ref{fig:range_RMSE_NLOS_LOS}, the impact of introducing a LOS path is evaluated. In this scenario, the transmitter and receiver are positioned at  $[-200,~ 0]$~[m] and $[200,~ 0]$~[m], respectively. The target location is randomly selected from the region defined by $x_{\min} = y_{\min} =  -200$ [m],  and $x_{\max} = 200$ [m], $y_{\max} = -100$ [m].  The RMSE of the proposed approach after LOS path cancellation is plotted against the NLOS-to-LOS power ratio $\abs{\alpha_{\text{NLOS}}}^2/\abs{\alpha_{\text{LOS}}}^2$, for pilot patterns $(n_p, m_p)\in\{(2, 1), (2, 2)\}$. The results reveal that as the NLOS path becomes relatively stronger, the RMSE curves approach the CRBs, indicating improved estimation accuracy. However, beyond a certain threshold, further strengthening of the NLOS component causes imperfections in the LOS path removal process. This degrades both range and velocity estimation performance, as reflected by a noticeable divergence between the RMSE and CRB curves in Fig.~\ref{fig:range_RMSE_NLOS_LOS}.
	
	\section{Concluding Remarks}
	In this work, we have demonstrated that the ISI-free sensing range in bi-static ISAC configurations can be effectively extended using a sliding window-based approach. Since the sensing receiver does not have access to the full transmitted signal, it leverages pilot symbols embedded in the time-frequency grid for target detection and parameter estimation. The proposed framework addresses both LOS presence and blockage scenarios, and it is shown that the resulting receiver structures achieve performance that closely aligns with the CRBs for both range and velocity estimation. Future research directions include extending the framework to incorporate AoA and AoD estimation, as well as generalizing the approach to multi-static ISAC systems, which holds significant promise for enhanced spatial sensing capabilities.
	\vspace{-0.3cm}
	\bibliographystyle{IEEEtran}
	\bibliography{bibfile}
	
\end{document}

%% file: commands.tex
\usepackage{pgfplots} 
\usepackage{comment}
\usepackage{amsfonts}
\usepackage{textcomp}
\usepackage[T1]{fontenc}

\def\BibTeX{{\rm B\kern-.05em{\sc i\kern-.025em b}\kern-.08em
		T\kern-.1667em\lower.7ex\hbox{E}\kern-.125emX}}
\usepackage{amsmath,amssymb,multirow}
\usepackage{setspace}
\usepackage{listings}
\usepackage{cite}
\usepackage{acronym}
\usepackage{mathtools}
\usepackage{subfigure}
\usepackage{algorithm,algorithmic}
\usepackage{bm}

\pgfplotsset{
	tick label style={font=\small},
	label style={font=\small},
	legend style={font=\small}
}

\DeclareMathOperator*{\argmax}{arg\,max}

\newcommand\abs[1]{\big|#1\big|}

\newcommand\norm[1]{\left\lVert #1\right\rVert}

\newcommand{\Tcp}{T_{\mathrm{cp}}}
\newcommand{\Td}{T_{\mathrm{d}}}
\newcommand{\Ts}{T_{\mathrm{s}}}
\newcommand{\Tsym}{T_{\mathrm{sym}}}
\newcommand{\Nsym}{N_{\mathrm{sam}}}

\newcommand{\Ncp}{N_{\mathrm{cp}}}

\newcommand{\fbt}{\mathbf{f}_{T}}
\newcommand{\fbr}{\mathbf{f}_{R}}
\newcommand{\Wb}{\mathbf{W}}

\newcommand{\hermit}{\mathsf{H}}

\newcommand{\HLS}{\widehat{H}^{\text{LS}}}

\newcommand{\ab}{\boldsymbol{a}}
\newcommand{\pb}{\boldsymbol{p}}
\newcommand{\vb}{\boldsymbol{v}}
\newcommand{\pbtx}{\boldsymbol{p}_{\text{tx}}}
\newcommand{\pbrx}{\boldsymbol{p}_{\text{rx}}}

\newcommand{\Rbis}{R_{\text{bis}}}
\newcommand{\Rmax}{R_{\text{max}}}
\newcommand{\vbis}{v_{\text{bis}}}

\newcommand{\Rhatbis}{\widehat{R}_{\text{bis}}}
	
\newcommand{\Nsc}{N_{\text{sc}}}	
\newcommand{\Msym}{M_{\text{sym}}}

\makeatletter \renewcommand\d[1]{\ensuremath{%
		\;\mathrm{d}#1\@ifnextchar\d{\!}{}}}
\makeatother

\makeatletter
\newcommand*\rel@kern[1]{\kern#1\dimexpr\macc@kerna}
\newcommand*\widebar[1]{%
	\begingroup
	\def\mathaccent##1##2{%
		\rel@kern{0.8}%
		\overline{\rel@kern{-0.8}\macc@nucleus\rel@kern{0.2}}%
		\rel@kern{-0.2}%
	}%
	\macc@depth\@ne
	\let\math@bgroup\@empty \let\math@egroup\macc@set@skewchar
	\mathsurround\z@ \frozen@everymath{\mathgroup\macc@group\relax}%
	\macc@set@skewchar\relax
	\let\mathaccentV\macc@nested@a
	\macc@nested@a\relax111{#1}%
	\endgroup
}
\makeatother

\usepackage{amsthm}

\theoremstyle{remark}

\newtheoremstyle{mytheoremstyle} 
{\topsep}                    
{\topsep}                    
{\upshape}                   
{.5em}                           
{\itshape}                   
{.}                          
{.5em}                       
{}  

\theoremstyle{mytheoremstyle}

\newtheoremstyle{iremark}
{\topsep}   
{\topsep}   
{\upshape}  
{0.2in}       
{\itshape}  
{.}         
{5pt plus 1pt minus 1pt} 
{\thmname{#1}\thmnumber{ \itshape#2}\thmnote{ (#3)}} 

\theoremstyle{plain}

%% file: main.bbl
\begin{thebibliography}{10}
\providecommand{\url}[1]{#1}
\csname url@samestyle\endcsname
\providecommand{\newblock}{\relax}
\providecommand{\bibinfo}[2]{#2}
\providecommand{\BIBentrySTDinterwordspacing}{\spaceskip=0pt\relax}
\providecommand{\BIBentryALTinterwordstretchfactor}{4}
\providecommand{\BIBentryALTinterwordspacing}{\spaceskip=\fontdimen2\font plus
\BIBentryALTinterwordstretchfactor\fontdimen3\font minus
  \fontdimen4\font\relax}
\providecommand{\BIBforeignlanguage}[2]{{%
\expandafter\ifx\csname l@#1\endcsname\relax
\typeout{** WARNING: IEEEtran.bst: No hyphenation pattern has been}%
\typeout{** loaded for the language `#1'. Using the pattern for}%
\typeout{** the default language instead.}%
\else
\language=\csname l@#1\endcsname
\fi
#2}}
\providecommand{\BIBdecl}{\relax}
\BIBdecl

\bibitem{Gonzalez2024ISAC}
N.~Gonzalez-Prelcic, M.~Furkan~Keskin, O.~Kaltiokallio, M.~Valkama, D.~Dardari,
  X.~Shen, Y.~Shen, M.~Bayraktar, and H.~Wymeersch, ``The integrated sensing
  and communication revolution for 6g: Vision, techniques, and applications,''
  \emph{Proceedings of the IEEE}, vol. 112, no.~7, pp. 676--723, 2024.

\bibitem{Furkan25Holistic}
M.~F. Keskin, M.~M. Mojahedian, J.~O. Lacruz, C.~Marcus, O.~Eriksson,
  A.~Giorgetti, J.~Widmer, and H.~Wymeersch, ``Fundamental trade-offs in
  monostatic isac: A holistic investigation towards 6g,'' \emph{IEEE
  Transactions on Wireless Communications}, pp. 1--1, 2025.

\bibitem{Swindlehurst2024Joint3D}
Z.~Xiao, R.~Liu, M.~Li, Q.~Liu, and A.~L. Swindlehurst, ``A novel joint
  angle-range-velocity estimation method for mimo-ofdm isac systems,''
  \emph{IEEE Transactions on Signal Processing}, vol.~72, pp. 3805--3818, 2024.

\bibitem{22_Pucci_SystemLevelAnalysis}
L.~Pucci, E.~Paolini, and A.~Giorgetti, ``System-level analysis of joint
  sensing and communication based on 5g new radio,'' \emph{IEEE Journal on
  Selected Areas in Communications}, vol.~40, no.~7, pp. 2043--2055, 2022.

\bibitem{Ozturk2025bistatic}
\BIBentryALTinterwordspacing
C.~Ozturk and C.~Goken, ``Impact of the pilot design for ofdm based bi-static
  integrated sensing and communication system,'' 2025. [Online]. Available:
  \url{https://arxiv.org/abs/2503.20288}
\BIBentrySTDinterwordspacing

\bibitem{bacchielli2024bistaticsensingthzfrequencies}
\BIBentryALTinterwordspacing
T.~Bacchielli, L.~Pucci, D.~Dardari, and A.~Giorgetti, ``Bistatic sensing at
  thz frequencies via a two-stage ultra-wideband mimo-ofdm system,'' 2024.
  [Online]. Available: \url{https://arxiv.org/abs/2405.17990}
\BIBentrySTDinterwordspacing

\bibitem{Pucci2022_Bistatic}
L.~Pucci, E.~Matricardi, E.~Paolini, W.~Xu, and A.~Giorgetti, ``Performance
  analysis of a bistatic joint sensing and communication system,'' in
  \emph{2022 IEEE International Conference on Communications Workshops (ICC
  Workshops)}, 2022, pp. 73--78.

\bibitem{brunner2024bistaticofdmbasedisacovertheair}
\BIBentryALTinterwordspacing
D.~Brunner, L.~G. de~Oliveira, C.~Muth, S.~Mandelli, M.~Henninger, A.~Diewald,
  Y.~Li, M.~B. Alabd, L.~Schmalen, T.~Zwick, and B.~Nuss, ``Bistatic ofdm-based
  isac with over-the-air synchronization: System concept and performance
  analysis,'' 2024. [Online]. Available: \url{https://arxiv.org/abs/2405.04962}
\BIBentrySTDinterwordspacing

\bibitem{Natajara2024BistaticRadar}
N.~K. Nataraja, S.~Sharma, K.~Ali, F.~Bai, R.~Wang, and A.~F. Molisch,
  ``Integrated sensing and communication (isac) for vehicles: Bistatic radar
  with 5g-nr signals,'' \emph{IEEE Transactions on Vehicular Technology}, pp.
  1--16, 2024.

\bibitem{Cho24MILCOM}
H.~Cho, S.~Yoo, B.~C. Jung, and J.~Kang, ``Enhancing battlefield awareness: An
  aerial ris-assisted isac system with deep reinforcement learning,'' in
  \emph{MILCOM 2024 - 2024 IEEE Military Communications Conference (MILCOM)},
  2024, pp. 469--474.

\bibitem{Sharma22MILCOM}
S.~Sharma and V.~Koivunen, ``Multicarrier ds-cdma based integrated sensing and
  communication waveform designs,'' in \emph{MILCOM 2022 - 2022 IEEE Military
  Communications Conference (MILCOM)}, 2022, pp. 95--101.

\bibitem{braun2014ofdm}
K.~M. Braun, ``Ofdm radar algorithms in mobile communication networks,'' Ph.D.
  dissertation, Karlsruhe, Karlsruher Institut f{\"u}r Technologie (KIT),
  Diss., 2014, 2014.

\bibitem{Keskin2021ICIFoe}
M.~F. Keskin, H.~Wymeersch, and V.~Koivunen, ``Mimo-ofdm joint
  radar-communications: Is ici friend or foe?'' \emph{IEEE Journal of Selected
  Topics in Signal Processing}, vol.~15, no.~6, pp. 1393--1408, 2021.

\bibitem{Tang24CP}
A.~Tang, Q.~Zhao, X.~Wang, and W.~Qu, ``Isi-resistant reference signal design
  and processing for ofdm integrated communications and long-range radar
  sensing,'' \emph{IEEE Communications Letters}, vol.~28, no.~6, pp.
  1322--1326, 2024.

\bibitem{Wang25CP}
L.~Wang, Z.~Wei, X.~Chen, and Z.~Feng, ``Coherent compensation-based sensing
  for long-range targets in integrated sensing and communication system,''
  \emph{IEEE Transactions on Vehicular Technology}, pp. 1--15, 2025.

\bibitem{Jiang25CP}
Q.~Jiang, X.~Sun, D.~Wang, C.~Pan, and J.~Wang, ``Scalable long-distance isac
  signal design for ofdm systems with theoretical analysis and practical
  validation,'' \emph{IEEE Wireless Communications Letters}, pp. 1--1, 2025.

\bibitem{xu2025CP}
\BIBentryALTinterwordspacing
X.~Xu, Z.~Zhou, and Y.~Zeng, ``How does cp length affect the sensing range for
  ofdm-isac?'' 2025. [Online]. Available:
  \url{https://arxiv.org/abs/2503.08062}
\BIBentrySTDinterwordspacing

\bibitem{willis2005bistatic}
N.~J. Willis, ``Bistatic radar,'' \emph{Scitech Publishing Inc. google
  scholar}, vol.~2, pp. 604--612, 2005.

\end{thebibliography}
